\documentclass[aps,twocolumn,showpacs,footinbib]{revtex4}%
\usepackage[latin9]{inputenc}
\usepackage{amssymb}
\usepackage{amsmath}
\usepackage{amscd}
\usepackage{graphicx}
\usepackage{subfigure}
\usepackage{float}
\begin{document}
\draft
\title{Manifestation of nonclassical Berry phase of an electromagnetic field in
atomic Ramsey interference}
\author{Shi-Biao Zheng}
\address{Department of Physics\\
Fuzhou University\\
Fuzhou 350002, P. R. China}
\date{\today }

\begin{abstract}
The Berry phase acquired by an electromagnetic field undergoing an adiabatic
and cyclic evolution in phase space is a purely quantum-mechanical effect of
the field. However, this phase is usually accompanied by a dynamical
contribution and can not be manifested in any light-beam interference
experiment because it is independent of the field state. We here show that
such a phase can be produced using an atom coupled to a quantized field and
driven by a slowly changing classical field, and it is manifested in the
atomic Ramsey interference oscillations. We also show how this effect may be
applied to one-step implementation of multi-qubit geometric phase gates,
which is impossible by previous geometric methods. The effects of
dissipation and fluctuations in the parameters of the pump field on the
Berry phase and visibility of the Ramsey interference fringes are analyzed.
\end{abstract}

\maketitle

\vskip 0.5cm

\narrowtext

\section{INTRODUCTION}

When the Hamiltonian of a quantum system, depending on a set of parameters,
is adiabatically changed along a closed curve in parameter space, then the
quantum system in an eigenstate of the Hamiltonian will acquire a purely
geometric phase in addition to the usual dynamical phase [1]. Compared with
the dynamic phase, the Berry phase is given by a circuit integral in
parameter space and is independent of energy and time. During the past few
decades, Berry phase has been the subject of a variety of theoretical and
experimental investigations. Besides the fundamental interest, Berry phase
has many important applications, ranging from optics and molecular physics
to quantum computation by geometric means [2-6]. Since the Berry phase only
depends upon the geometry of the evolution path, it is robust against
fluctuation perturbations that affect the dynamical phase [7]. This feature
makes quantum logic gates based on geometric phases have potential
fault-tolerance in the presence of noise perturbation. The geometric phase
has been generalized to the case of nonadiabatic, noncyclic, and nonunitary
evolution of a quantum system [8,9].

The Berry phase and its robustness against noise perturbations has been
experimentally tested in various two-state systems [10-14]. Optical
experiments have been performed to observe Berry phase of light beams
[15-17] that can be understood as a classical effect following the Maxwell
equations [18]. The observation of this effect at the single-photon level
has also been reported [19], but the Berry phase without classical origin
has not been directly measured for any quantum harmonic oscillator in
continuous-variable (infinite-dimensional) states. The cyclic and adiabatic
displacement in phase space is the simplest quantum-mechanical
transformation that can produce a nonclassical Berry phase for a
continuous-variable system. Unfortunately, no scheme has been proposed for
realizing such transformation in a realistic physical system without
introducing the dynamical phase. Furthermore, the Berry phase acquired
through such a transformation cannot be manifested in any optical
interference experiment. This is due to the fact that the interference of
light fields is fundamentally different from that of particles. It is the
relative phase of states that manifests in the latter case, while it is the
relative phase of the electric amplitudes in the former case. Thus, the
geometric phase measured in a light-beam interference experiment is the
Hannay angle rather than the Berry phase [20]. For a cyclic displacement
evolution, the acquired Berry phase is independent of the field state and
the Hannay angle is zero.

Here we show that the Berry phase of a quantized field, associated with a
cyclic and adiabatic displacement in phase space, can be produced and
measured with an atomic quantum bit (qubit) that is coupled to the quantized
field and driven by a classical pump field. By means of variation of the
parameters of the pump field the quantized field undergoes an adiabatic and
cyclic evolution in phase space, conditional upon the state of the qubit.
The two qubit states correspond to two evolution paths in the Hilbert space,
one correlated with the adiabatic displacement of the quantized field and
the other correlated with free evolution. The Berry phase of the quantized
field is manifested in the interference of the atomic qubit, other than in
the interference of the field itself. As far as we know, our proposal is the
first realistic one for directly measuring the Berry's adiabatic geometric
phase for infinite-dimensional field states. When multiple qubits are
involved, the acquired Berry phase leads to one-step implementation of
important geometric quantum gates, which can not be achieved by previous
methods. The effects of dissipation and the fluctuations in the parameters
of the pump field on this phase and the atomic coherence are analyzed. The
required qubit-boson coupling can be realized in cavity or circuit QED
systems with atomic or superconducting qubits coupled to a resonator.

The paper is organized as follows. In Sec.2 we propose a scheme for
measuring the Berry phase of a quantized field using the Ramsey interference
of an atomic qubit coupled to the quantized field and driven by an external
field. We show that the acquired Berry phase can be used for one-step
implementation of $n$-qubit quantum phase gates in Sec.3. In Sec.4 we
investigate the geometric phase and visibility of the Ramsey interference
fringes with the field decay and atomic spontaneous emission being included.
Sec. 5 sees an analysis of the effect of the fluctuations in the parameters
of the pump field on the Berry phase and atomic coherence. Finally we
dissuss the physical implementation of the model in microwave cavity QED and
generalization of the ideas to the ion trap system, and we present a summary
of our results in Sec. 6.

\section{MEASUREMENT OF THE BERRY PHASE OF THE QUANTIZED FIELD}

We first consider a two-level atom resonantly interacting with a single-mode
quantized electromagnetic field and driven by a classical field. The
dynamics of the whole system is described by the driven Jaynes-Cummings
model [21]. We will label the upper and lower states of the two-level atom
as $\left| e\right\rangle $ and $\left| g\right\rangle $ and describe its
dynamics in terms of the Pauli operators $\sigma _z$, $\sigma ^{\pm
}=(\sigma _x\pm i\sigma _y)/2$. Then the Hamiltonian, in the interaction
picture, is (assuming $\hbar =1$)
\begin{equation}
H=\lambda (a^{\dagger }\sigma ^{-}+a\sigma ^{+})+\Omega (\sigma
^{-}e^{-i\phi }+\sigma ^{+}e^{i\phi }),
\end{equation}
where $a^{\dagger }$ and $a$ are the creation and annihilation operators for
the quantized field, $\lambda $ is the atom-cavity coupling strength, and $%
\Omega $ and $\phi $ are the Rabi frequency and phase of the pump field. The
first part of the Hamiltonian, describing the coherent energy exchange
between the two-level atom and the quantized field, corresponds to the
normal Jaynes-Cummings model [22]. The second part represents the coupling
between the atomic transition $\left| g\right\rangle \leftrightarrow \left|
e\right\rangle $ and the classical field. The eigenenergies of the driven
Jaynes-Cummings model are the same as those of the normal Jaynes-Cummings
model, while the field parts of the eigenstates are displaced in phase space
with the displacement parameter determined by the Rabi frequency and phase
of the pump field [21]. This allows one to adiabatically drive the quantized
field to undergo a cyclic displacement evolution by slowly changing the
parameters of the pump field. The Berry phase gained this way is determined
by the area enclosed by the phase-space displacement trajectory, which is
independent of the state of the system. To illustrate the idea clearly, we
first consider the evolution of the dark state (the eigenstate with zero
eigenenergy) of the Hamiltonian $H$. Such a state is $\left| \psi _0(\alpha
)\right\rangle =\left| g\right\rangle \left| \alpha \right\rangle $, where $%
\left| \alpha \right\rangle $ is the coherent state of the quantized field
with the parameter $\alpha =-\Omega e^{i\phi }/\lambda $. Adiabatic
following of the dark state renders the phase of the quantized field to be
opposite to that of the pump field. When the parameters of the Hamiltonian
complete a cyclic evolution after the duration $T$, the dark state describes
a displacement loop of the quantized field and acquires a Berry phase
\begin{eqnarray}
\beta &=&i\int_0^Tdt\left\langle \psi _0(\alpha )\right| \frac d{dt}\left|
\psi _0(\alpha )\right\rangle \\
\ &=&\frac i2\oint (\alpha ^{*}d\alpha -\alpha d\alpha ^{*})=\pm 2A,
\nonumber
\end{eqnarray}
where $A$ is the area enclosed by the phase-space loop. The $\pm $
sign depends on whether the sense of rotation is clockwise or
counter-clockwise. Suppose that the Rabi frequency $\Omega $ of the
pump field is kept constant and the phase $\phi $, serving as the
control parameter, is slowly varied from $0$ to $2\pi $. Then the
state of the quantized field is moved around a circle with radius
$\left| \alpha \right| $ in phase space and the acquired Berry phase
is given by $-2\pi \left| \alpha \right| ^2$. During the adiabatic
evolution, the atom remains in the lower state $\left|
g\right\rangle $ and the Berry phase completely originates from the
cyclic quantum-mechanical displacement of the quantized field
induced by the adiabatic variation of the Hamiltonian. This phase is
significantly different from the nonadiabatic geometric phase
produced by a coherent displacement force [23] in that the latter
involves a nonzero dynamical component that is proportional to the
total phase and thus cannot be removed [8, 24]. We note that the
displacement evolution path of the quantized field is not affected
if a term proportional to $\sigma _z$ is added to the Hamiltonian.
This implies that the acquired Berry phase is immune from the
fluctuation in the qubit transition frequency as compared with the
geometric phase of a qubit coupled only to a classical field.

In order to interpret the acquired Berry phase in the parameter space we
rewrite the Hamiltonian as $H=\lambda [(a^{\dagger }\sigma ^{-}+a\sigma
^{+})+{\bf B}\cdot {\bf \sigma }]$, where ${\bf \sigma =\{\sigma }_x$, ${\bf %
\sigma }_y$, ${\bf \sigma }_z{\bf \}}$ and ${\bf B}=(\Omega \cos \phi
/\lambda $, $\Omega \sin \phi /\lambda $, $0)$ is the dimensionless
effective vector field which acts as the control parameter. By cyclically
changing $\Omega $ and/or $\phi $ the Hamiltonian describes a closed path in
the two-dimensional parameter space $\left\{ {\bf B}\right\} $. When the
system is initially in the dark state of the Hamiltonian, the state of the
system will follow the effective field and after a closed cycle it gains a
purely geometric phase proportional to the area enclosed by the circuit
traversed by the effective field. Note that, although the two-level system
is coupled to the slowly changing effective vector field, its state is not
varied when the system is initially in the dark state because the transition
paths induced by the two fields with opposite phases interfere
destructively. It is the quantized field whose state adiabatically follows
the effective field ${\bf B}$ and eventually acquires the Berry phase which
depends upon the global property of the evolution path in the parameter
space of the effective field.

The eigenstates of the Hamiltonian $H$ with nonzero eigenvalues are $\left|
\psi _{n+1,\pm }(\alpha )\right\rangle =D(\alpha )(\left| g\right\rangle
\left| n+1\right\rangle \pm \left| e\right\rangle \left| n\right\rangle )/%
\sqrt{2}$, where $D(\alpha )=e^{\alpha a^{\dagger }-\alpha ^{*}a}$ is the
displacement operator, and $n+1$ is the quantum number of the displaced
field state. After a complete evolution cycle all the eigenstates gain the
same Berry phase $\beta $, which makes it impossible to observe the Berry
phase in an interference experiment by initially preparing a superposition
of different eigenstates as no relative geometric phase can be obtained.
Neither can this phase be observed in optical interference experiments
because it does not depend on the state of the quantized field. In the
following, we show such a phase can be manisfested in the interference
between the probability amplitudes associated with the atomic state $\left|
g\right\rangle $ and an auxiliary state $\left| f\right\rangle $ which is
neither coupled to the pump field nor to the quantized field mode. The atom
is initially driven to the superposition state $\frac 1{\sqrt{2}}(\left|
g\right\rangle +\left| f\right\rangle )$ from $\left| f\right\rangle $ using
a classical pulse, which is analogous to the splitting of the photon beam at
the first beam splitter of a Mach-Zehnder interferometer. The quantized
field, initially prepared in the coherent state $\left| -\Omega /\lambda
\right\rangle $, acts as the dephasing element in one arm of the
interferometer. After adiabatically dragging the Hamiltonian of Eq. (1)
along a closed loop the quantized field undergoes a conditional cyclic
displacement in phase space and introduces a relative phase between the two
atomic states $\left| g\right\rangle $ and $\left| f\right\rangle $, leading
to the superposition state $\frac 1{\sqrt{2}}(e^{i\beta }\left|
g\right\rangle +\left| f\right\rangle )$. This implies that the Berry phase
acquired by the quantized field is encoded in the probability amplitude for
finding the atom in the state $\left| g\right\rangle $. The interference
between the probability amplitudes of the two superposed atomic states can
be achieved through the transformations $\left| g\right\rangle
\longrightarrow \frac 1{\sqrt{2}}(\left| g\right\rangle -\left|
f\right\rangle )$ and $\left| f\right\rangle \longrightarrow \frac 1{\sqrt{2}%
}(\left| f\right\rangle +\left| g\right\rangle )$, which are analogous to
the recombination of the photon beams at the second beam splitter of the
Mach-Zehnder interferometer. Finally, the atomic state is measured. The
probabilities of finding the atom in the states $\left| g\right\rangle $ and
$\left| f\right\rangle $ are given by

\begin{equation}
P_{g,f}=\frac 12(1\pm \cos \beta ),
\end{equation}
which is independent of the atom-field interaction time. By varying the
Berry phase $\beta $ the probability of finding the atom in a definite state
exhibits Ramsey interference fringes. Therefore, the Berry phase of light
field is manisfested in the atomic Ramsey interference. With the choice $%
\Omega =\lambda /\sqrt{2}$ and $\left| \alpha \right| =1/\sqrt{2}$, an
adiabatic and cyclic evolution from $\phi =0$ to $\phi =2\pi $ achieves the
Berry phase $\beta =-\pi $. Due to the presence of the Berry phase the atom
is finally in the state $\left| g\right\rangle $. On the other hand, if no
Berry phase is present, the atom is finally in the state $\left|
f\right\rangle $. Therefore, the detection of the final state of the atom
unambigually distinguishes whether the Berry phase is present or not. It is
important to note that even if the quantized field is in a macroscopic
coherent state with $\left| \alpha \right| \gg 1$ the adiabatic and cyclic
evolution of this field can be engineered by changing the Hamiltonian around
a suitable circuit in parameter space, enabling the nonclassical Berry phase
of light to be tested at the macroscpic level.

A modification of the interferometer can be applied to measure the Berry
phase for any initial field state, for example, a thermal state. If the
system is not initially in the dark state a purely geometric phase can be
observed by applying a phase kick $-\sigma _z$ to the atom at time $T/2$.
The product of the atomic state $\left| g\right\rangle $ with any displaced
number state $D(\alpha )\left| n+1\right\rangle $ can be expressed as a
superposition of the two eigenstates $\left| \psi _{n+1,+}(\alpha
)\right\rangle $ and $\left| \psi _{n+1,-}(\alpha )\right\rangle $. During
the adiabatic evolution these two eigenstates acquire opposite dynamical
phases. The phase kick inverts these two eigenstates, effectively inverting
the dynamical phases accumulated from time $0$ to $T/2$. At time $T$ the
system completes a nontrivial cyclic evolution and the dynamical phase is
completely canceled. The whole procedure results in a purely geometric phase
$\beta $, which is determined by the area of the circuit followed by the
effective field ${\bf B}$ in parameter space and independent of the initial
field state. This implies that the quantized field does not need to be
prepared in a specific state since any field state can be expressed in terms
of coherent state $\left| \alpha \right\rangle $ and displaced number states
$D(\alpha )\left| n+1\right\rangle $. After the cyclic evolution all
components acquire the same geometric phase $\beta $ and the dynamical
phases associated with all displaced number states $D(\alpha )\left|
n+1\right\rangle $ are removed via the application of the atomic phase kick.
The atomic interferometer can also be used to measure the geometric phase
for the quantized field undergoing a noncyclic evolution. In this case the
geometric phase can be expressed as the minus double of the area enclosed by
the displacement trajectory and the straight line (the null phase curve)
connecting the starting and ending points in phase space [25]. The geometric
phase is directly related to the shift of the Ramsey interference fringes
[26].

\section{ONE-STEP IMPLEMENTATION OF MULTI-QUBIT PHASE GATES}

Besides being of fundamental interest, geometric phases can be applied to
design of quantum logic gates that have an intrinsic resilience against
noise perturbations. We note that the conditional phase gates for $n$ atomic
qubits can be produced via a single conditional adiabatic displacement of
the quantized field. The computational basis for each qubit is formed by the
two states $\left| g\right\rangle $ and $\left| f\right\rangle $. If the
transition $\left| g\right\rangle \leftrightarrow \left| e\right\rangle $ of
each qubit is resonantly coupled to a quantized field and driven by a pump
field, the interaction Hamiltonian is
\begin{equation}
H_n=\sum_{j=1}^n[\lambda _j(a^{\dagger }\sigma _j^{-}+a\sigma _j^{+})+\Omega
_j(e^{-i\phi }\sigma _j^{-}+e^{i\phi }\sigma _j^{+})],
\end{equation}
where the subscript $j$ labels the qubits. Under the condition $\Omega
_1/\lambda _1=\Omega _2/\lambda _2=...=\Omega _n/\lambda _n=r$ the
Hamiltonian $H_n$ has dark states of the form $\left| \phi _a\right\rangle
\left| -re^{i\phi }\right\rangle $, where $\left| \phi _a\right\rangle $ can
be any computational basis state except $\left| f_1f_2...f_n\right\rangle $.
If the system is initially in the state $\left| \phi _a\right\rangle \left|
-r\right\rangle $, slow variation of the phase $\phi $ from $0$ to $2\pi $
produces a Berry phase $\beta $. On the other hand, the state $\left|
f_1f_2...f_n\right\rangle \left| -r\right\rangle $ is completely decoupled
from the Hamiltonian $H_n$. As the state $\left| f\right\rangle $ is not
coupled to the fields, transitions between degenerate dark states do not
occur. Then the evolution of the qubit system proceeds as $\left| \phi
_a\right\rangle \rightarrow e^{i\beta }\left| \phi _a\right\rangle $ and $%
\left| f_1f_2...f_n\right\rangle \rightarrow e^{i\beta }e^{-i\beta }\left|
f_1f_2...f_n\right\rangle $. Discarding the trivial common phase factor $%
e^{i\beta }$, this is equivalent to an $n$-qubit controlled phase gate
\begin{equation}
U_n=e^{-i\beta \left| f_1f_2...f_n\right\rangle \left\langle
f_1f_2...f_n\right| },
\end{equation}
in which if and only if all the qubits are in the state $\left|
f\right\rangle $ the system undergoes a phase shift $-\beta $. This gate is
essential to implementation of Grover's algorithms [27] and quantum Fourier
transform [28]. Though any quantum computational network can be decomposed
into a series of two- plus one-qubit logic gates, it may be extremely
complex to construct an $n$-qubit phase gate using these elementary gates as
the number of required operations increases exponentially with $n$. Thus,
direct realization of $n$-qubit phase gates would greatly simplify practical
implementation of certain quantum computational tasks. We note that the
nonadiabatic geometric means [23] can be directly used for implementation of
the $n$-qubit entangling gate $U_n^{^{\prime }}=\exp (i\theta J_z^2)$ where $%
J_z=\frac 12\sum_{j=1}^n(\left| f_j\right\rangle \left\langle f_j\right|
-\left| g_j\right\rangle \left\langle g_j\right| )$, but it does not allow
one-step implementation of the phase gate $U_n$.

Another important feature of the present gate operation is that it does not
require the qubit-resonator coupling strengths $\lambda _j$ to be identical.
Furthermore, the conditional phase shift is not affected even if the
transition frequencies of the qubits are different because the evolution of
the quantized field is not changed when we add to the Hamiltonian $H_n$ the
terms $\sum_{j=1}^n\delta _j\sigma _{z,j}$, where $\delta _j$ is detuning
between the transition frequency of qubit $j$ and the field frequency. The
tolerance to the nonuniformity of the qubits is important for the
solid-state implementation of quantum computation since the parameters of
artificial atomic qubits are usually not uniform. The geometric phase gate
can also be implemented with the quantized field being initially in any
state by applying the phase kick to all the atoms at time $T/2$.

\section{THE EFFECT OF DISSIPATION}

Now let us derive the geometric phase and the visibility of the Ramsey
interference fringes with the dissipation being included. The evolution of
the system is governed by the master equation
\begin{equation}
\stackrel{.}{\rho }=-i[H,\rho ]+{\cal L}_\gamma \rho +{\cal L}_\kappa \rho ,
\end{equation}
where
\begin{eqnarray}
{\cal L}_\gamma \rho &=&\frac \gamma 2(2a\rho a^{\dagger }-\rho
a^{\dagger }a-a^{\dagger }a\rho ),  \nonumber \cr {\cal L}_\kappa
\rho &=&\sum_j\sum_k\frac{\kappa _{j,k}}2(2S_{j,k}^{-}\rho
S_{j,k}^{+}-\rho
S_{j,k}^{+}S_{j,k}^{-}-S_{j,k}^{+}S_{j,k}^{-}\rho ),\\
\end{eqnarray}
$S_{j,k}^{+}=\left| j\right\rangle \left\langle k\right| $, $S^{-}=\left|
k\right\rangle \left\langle j\right| $, $\gamma $ is the decay rate for the
quantized field, and $\kappa _{j,k}$ is the rate for the atomic spontaneous
emission $\left| j\right\rangle \rightarrow \left| k\right\rangle $ ($\left|
k\right\rangle $ being the levels lower than $\left| j\right\rangle $). The
evolution of the system during the infinitesimal interval [$t$, $t+dt$] is
[29]
\begin{equation}
\rho (t)\rightarrow \rho (t+dt)=e^{{\cal L}_\kappa dt}e^{{\cal L}_\gamma dt}%
{\cal U}(t,dt)\rho (t),
\end{equation}
where
\begin{equation}
{\cal U}(t,dt)\rho (t)=U(t,dt)\rho (t)U^{\dagger }(t,dt),
\end{equation}
with $U(t,dt)$ being the unitary evolution operator governed by the slowly
changing Hamiltonian $H$ during [$t$, $t+dt$]. In the coherent state basis
the action of the superoperator $e^{{\cal L}_\gamma dt}$ on the elements of
the density matrix is given by [30]
\begin{equation}
e^{{\cal L}_\gamma dt}\left| \alpha _1\right\rangle \left\langle \alpha
_2\right| =\left\langle \alpha _2\right| \left. \alpha _1\right\rangle
^{\gamma dt}\left| \alpha _1e^{-\gamma dt}\right\rangle \left\langle \alpha
_2e^{-\gamma dt}\right| .
\end{equation}
We here have discarded the atomic part in the density matrix element, which
is not affected by $e^{{\cal L}_\gamma dt}$. The action of the superoperator
$e^{{\cal L}_\kappa dt}$ is given by
\begin{eqnarray}
e^{{\cal L}_\kappa dt}\left| g\right\rangle \left\langle g\right|
&=&e^{-\kappa _gdt}\left| g\right\rangle \left\langle g\right| +\sum_j\frac{%
\kappa _{g,j}}{\kappa _g}(1-e^{-\kappa _gdt})\left| j\right\rangle
\left\langle j\right| ,  \nonumber \\
e^{{\cal L}_\kappa dt}\left| f\right\rangle \left\langle f\right|
&=&e^{-\kappa _fdt}\left| f\right\rangle \left\langle f\right| +\sum_k\frac{%
\kappa _{f,j}}{\kappa _f}(1-e^{-\kappa _fdt})\left| k\right\rangle
\left\langle k\right| ,  \nonumber \\
e^{{\cal L}_\kappa dt}\left| g\right\rangle \left\langle f\right|
&=&e^{-(\kappa _g+\kappa _f)dt/2}\left| g\right\rangle \left\langle f\right|
,  \nonumber \\
e^{{\cal L}_\kappa dt}\left| f\right\rangle \left\langle g\right|
&=&e^{-(\kappa _g+\kappa _f)dt/2}\left| f\right\rangle \left\langle g\right|
,
\end{eqnarray}
where $\kappa _g=\sum_j\kappa _{g,j}$, $\kappa _f=\sum_k\kappa _{f,k}$, and $%
\left| j\right\rangle $ and $\left| k\right\rangle $ are levels lower than $%
\left| g\right\rangle $ and $\left| f\right\rangle $, respectively. The
dissipation makes the system deviate from the dark state of the Hamiltonian
and acquire a dynamical phase, which can be removed by frequently performing
the atomic phase kick $-\sigma _z$ during the course of the evolution of the
system ($M$ times with $M\gg 1$). When $M/T\gg \gamma $, $\kappa _e$, and $%
\kappa _g$, the dynamical effect is canceled before dissipation can affect
it, where $\kappa _e=\sum_j\kappa _{e,j}$. When the dynamical effect is
removed the action of the unitary evolution operator on the off-diagonal
matrix elements is
\begin{eqnarray}
&&\ \ \ U(t,dt)\left| g\right\rangle \left\langle f\right| \otimes \left|
\alpha _1\right\rangle \left\langle \alpha _2\right| U^{\dagger }(t,dt)
\nonumber \\
\ &=&e^{-iIm(\alpha _1^{*}d\alpha _0)}\left| g\right\rangle \left\langle
f\right| \otimes \left| \alpha _1+d\alpha _0\right\rangle \left\langle
\alpha _2\right| ,
\end{eqnarray}
where $\alpha _0=-\Omega e^{i\phi }/\lambda $.

Suppose that the Rabi frequency $\Omega $ of the pump field is kept constant
and the phase $\phi $ is slowly varied from $0$ to $2\pi $ with the constant
angular velocity $\omega =2\pi /T$ during the interval [$0$, $T$]. For the
initial state $\frac 1{\sqrt{2}}(\left| g\right\rangle +\left|
f\right\rangle )\left| -\Omega /\lambda \right\rangle $, the state of the
system at time $T$ is
\begin{eqnarray}
\rho (T) =\frac 12(e^{-\kappa _gT}\left| g\right\rangle \left\langle
g\right| \otimes \left| \alpha _g\right\rangle \left\langle \alpha
_g\right| +e^{-\kappa _fT}\left| f\right\rangle \left\langle
f\right| \cr \otimes \left| \alpha _f\right\rangle \left\langle
\alpha _f\right| \ \ +e^{-\Gamma +i\theta }\left| g\right\rangle
\left\langle f\right| \otimes \left| \alpha _g\right\rangle
\left\langle \alpha _f\right|\cr +e^{-\Gamma -i\theta }\left|
f\right\rangle \left\langle g\right| \otimes \left| \alpha
_f\right\rangle \left\langle \alpha _g\right| )   \ \ \cr+\frac
12\sum_j\left| j\right\rangle \left\langle j\right| \otimes
\int_0^T\kappa _{g,j}e^{-\kappa _gt}\left| \alpha _g^{^{\prime
}}\right\rangle \left\langle \alpha _g^{^{\prime }}\right| dt
\nonumber \cr \ \ +\frac 12\sum_k\frac{\kappa _{f,k}}{\kappa
_f}(1-e^{-\kappa _fT})\left| k\right\rangle \left\langle k\right|
\otimes \left| \alpha _f\right\rangle \left\langle \alpha _f\right|
,\\
\end{eqnarray}
where
\begin{eqnarray}
\alpha _g &=&\frac r{\gamma +i\omega }(i\omega +\gamma e^{-\gamma T}),
\nonumber \\
\alpha _g^{^{\prime }} &=&\frac r{\gamma +i\omega }(i\omega e^{i\omega
t}+\gamma e^{-\gamma t})e^{-(T-t)},  \nonumber \\
\alpha _f &=&re^{-\gamma T},  \nonumber \\
\Gamma &=&(\kappa _g+\kappa _f)T/2+\frac{r^2\omega ^2}{\gamma ^2+\omega ^2}[%
\frac 12\gamma T+\frac 14(1-e^{-2\gamma T})],  \nonumber \cr
\theta &=&-r^2[2\pi -\frac{\omega \gamma ^2T}{\gamma ^2+\omega ^2}-\frac{%
\omega \gamma (\omega ^2-2\gamma ^2)}{(\gamma ^2+\omega
^2)^2}(1-e^{-2\gamma T})],\\
\end{eqnarray}
$r=\Omega /\lambda $. We here have discarded the events that the atom
spontaneously emits two or more photons. After the transformations $\left|
g\right\rangle \longrightarrow \frac 1{\sqrt{2}}(\left| g\right\rangle
-\left| f\right\rangle )$ and $\left| f\right\rangle \longrightarrow \frac 1{%
\sqrt{2}}(\left| f\right\rangle +\left| g\right\rangle )$, the probabilities
of finding the atom in the states $\left| g\right\rangle $ and $\left|
f\right\rangle $ are given by
\begin{equation}
P_{g,f}=\frac 14[e^{-\kappa _gT}+e^{-\kappa _fT}+\frac{\kappa _{g,f}}{\kappa
_g}(1-e^{-\kappa _gT})\pm 2\nu \cos \beta ],
\end{equation}
where
\begin{eqnarray}
\nu &=&\exp [-\Gamma +\frac{r^2\omega ^2}{\gamma ^2+\omega ^2}(e^{-\gamma T}-%
\frac 12e^{-2\gamma T}-\frac 12)],  \nonumber \\
\beta &=&\theta -\frac{r^2\omega \gamma }{\gamma ^2+\omega ^2}e^{-\gamma T}.
\end{eqnarray}
We here have assumed that the level $\left| f\right\rangle $ is lower than $%
\left| g\right\rangle $. Under the condition $\kappa _g$, $\kappa _f$, $%
\gamma \ll 1/T$, to first order correction the visibility of the
interference fringes and geometric phase are approximately $\nu \simeq
1-r^2\gamma T-\kappa _{g,f}T/2$ and $\beta \simeq -\left| \alpha _0\right|
^2(2\pi +\gamma /\omega )$. The spontaneous emissions $\left| g\right\rangle
\rightarrow \left| j\right\rangle $ ($j\neq f$) and $\left| f\right\rangle
\rightarrow \left| k\right\rangle $ decrease the probabilities of finding
the atom in the states $\left| g\right\rangle $ and $\left| f\right\rangle $%
, but do not affect the visibility of the interference fringes since the sum
of the diagonal matrix elements and the sum of the off-diagonal matrix
elements of the reduced density oprator of the atom are shrunk by the same
factor due to these spontaneous emissions. It should be noted that the
second order spontaneous emissions $\left| j\right\rangle \rightarrow \left|
k\right\rangle $ following $\left| g\right\rangle \rightarrow \left|
j\right\rangle $ ($j\neq f$) or $\left| f\right\rangle \rightarrow \left|
j\right\rangle $ affect neither the probabilities of finding the atom in the
states $\left| g\right\rangle $ and $\left| f\right\rangle $ nor the
visibility of the interference fringes. Due to the second order spontaneous
emission $\left| f\right\rangle \rightarrow \left| k\right\rangle $
following $\left| g\right\rangle \rightarrow \left| f\right\rangle $ the
correction to the visibility of the interference fringes is on the order of $%
\kappa _{g,f}\kappa _fT^2$.

\section{THE EFFECT OF FLUCTUATIONS IN THE PARAMETERS OF THE PUMP FIELD}

We now analyze the effect of the fluctuations in the parameters of the pump
field on the Berry phase and atomic coherence. Set the fluctuations of the
Rabi frequency and phase of the pump field to be $\delta \Omega (t)$ and $%
\delta \phi (t)$. Due to the presence of the fluctuation noise, the dark
state of the system is $\left| g\right\rangle \left| \alpha (t)\right\rangle
$, where $\alpha (t)=-[r_0(t)+\delta r(t)]e^{i\phi (t)}$, $\phi (t)=\phi
_0(t)+\delta \phi (t)$, $r_0(t)=\Omega _0(t)/\lambda $, and $\delta
r(t)=\delta \Omega (t)/\lambda $. Here $\Omega _0(t)$ and $\phi _0(t)$ are
the Rabi frequency and phase of the pump field in the absence of noise. If
the system is initially in the dark state $\left| g\right\rangle \left|
\alpha (0)\right\rangle $ and the adiabatic condition is satisfied the final
state is $e^{i\theta (T)}\left| g\right\rangle \left| \alpha
(T)\right\rangle $, where $\theta (T)=-\int_{\phi (0)}^{\phi (T)}(r_0+\delta
r)^2d\phi $. To the first order correction the geometric phase is
\begin{equation}
\beta \simeq \beta _0-\int_0^T2r_0\delta r\stackrel{\cdot }{\phi }%
_0dt-\int_0^Tr_0^2\delta \stackrel{\cdot }{\phi }dt+r_0^2(0)\sin \delta \phi
(T),
\end{equation}
where $\beta _0$ is the Berry phase in the absence of noise. We here have
assumed $\delta r(0)=0$ and $\delta \phi (0)=0$. Without loss of the
generality, we again consider the case that $\Omega _0$ is kept constant and
$\phi _0$ undergoes an adiabatic cyclic evolution from $0$ to $2\pi $ with
the constant change rate $\stackrel{\cdot }{\phi }_0=2\pi /T$. Then the
expression reduces to $\beta \simeq \beta _0-\frac{4\pi }Tr_0\int_0^T\delta
rdt$. We here assume that the fluctuations are Gaussian and Markovian
processes with the bandwidth $\Gamma _\Omega $ ($\Gamma _\phi $) and
intensity $\sigma _\Omega ^2$ ($\sigma _\phi ^2$) for $\delta \Omega $ ($%
\delta \phi $). Under the condition $\Gamma _\Omega $, $\Gamma _\phi \ll
\lambda $, the pump field fluctuations are adiabatic with respect to the
Rabi frequency of the dressed atom-field system [7]. Consequently, the
geometric phase is Gaussian distributed with the mean value equal to the
noiseless Berry phase. Its variance is given by
\begin{equation}
\sigma _\beta ^2=16\pi \left| \beta _0\right| \frac{\sigma _\Omega ^2}{%
(\lambda \Gamma _\Omega T)^2}(\Gamma _\Omega T-1+e^{-\Gamma _\Omega T}).
\end{equation}
In the limit of the evolution being much slower than the fluctuation of the
Rabi frequency ($\Gamma _\Omega T\gg 1$), the variance approximates $\sigma
_\beta ^2=16\pi \left| \beta _0\right| \sigma _\Omega ^2/(\lambda ^2\Gamma
_\Omega T)$, which vanishes in the limit $(\Gamma _\Omega T)^{-1}\rightarrow
0$. When $\Gamma _\Omega T\ll 1$, the variance reduces to $\sigma _\beta
^2=8\pi \left| \beta _0\right| \sigma _\Omega ^2/\lambda ^2$. This implies
that the geometric phase is sensitive to slow fluctuations, with the
variance being path-dependent like the geometric phase itself, coinciding
with the geometric dephasing for a spin 1/2 in a slowly changing magnetic
field [7]. As long as the adiabatic condition is satisfied, the fluctuation
does not induce any correction to the dynamical phase as the spectrum of the
Hamiltonian does not depend on the parameters of the pump field. Thus the
dephasing of this system is different from that of the spin 1/2 to which the
main contribution has the dynamical origin [7].

Let us consider the effect of fluctuations in the classical control
parameters on the atomic Ramsey interference. In the presence of the noise,
after the conditional displacement evolution the system is in a mixed state,
whose density operator is given by
\begin{equation}
\rho =\int \left| \psi \right\rangle \left\langle \psi \right| P(\beta
)P(\delta r(T))P(\delta \phi (T))d\beta d\delta r(T)d\delta \phi (T),
\end{equation}
where
\begin{equation}
\left| \psi \right\rangle =\frac 1{\sqrt{2}}[e^{i\theta (T)}\left|
g\right\rangle \left| \alpha (T)\right\rangle +\left| f\right\rangle \left|
\alpha (0)\right\rangle )],
\end{equation}
$P(\beta )$, $P(\delta r(T))$, and $P(\delta \phi (T))$ are Gaussian
distribution functions for $\beta $, $\delta r(T)$, and $\delta \phi (T)$,
respectively. The coherence between the two atomic states is shrunk by a
factor $F=e^{-\sigma _\beta ^2/2-\sigma _\Omega ^2/(2\lambda ^2)-\left|
\beta _0\right| \sigma _\phi ^2/(4\pi )}$ due to random distributions in the
Berry phase and the noncyclic evolution of the field state correlated with
the atomic state $\left| g\right\rangle $.

Now we analyze how the classical noises affect the fidelity of the $n$-qubit
quantum gates. Set the initial state of the qubit system to be $c_1\left|
\psi _a\right\rangle +c_2$ $\left| f_1f_2...f_n\right\rangle $, where $%
\left| c_1\right| ^2+\left| c_2\right| ^2=1$ and $\left| \psi
_a\right\rangle $ can be any normalized superposition of the computational
basis states except $\left| f_1f_2...f_n\right\rangle $. Due to the
fluctuation perturbation the density operator of the whole system again has
the form of Eq. (19), where
\begin{equation}
\left| \psi \right\rangle =c_1e^{i\theta (T)}\left| \psi _a\right\rangle
\left| \alpha (T)\right\rangle +c_2\left| f_1f_2...f_n\right\rangle \left|
\alpha (0)\right\rangle .
\end{equation}
The infidelity caused by the fluctuations is $\epsilon =\left| c_1c_2\right|
^2(1-F)$. The result is valid even if these qubits are entangled with other
qubits not invovled in the gate operation.

\section{DISCUSSION AND CONCLUSION}

For potential physical implementation of the model, we consider microwave
cavity QED experiments with circular Rydberg atoms and a superconducting
millimeter-wave cavity, which has a remarkably long damping time $T_c=0.13$
s [31]. The states $\left| f\right\rangle $, $\left| g\right\rangle $, and $%
\left| e\right\rangle $ are the circular states with principal quantum
numbers 49, 50, and 51, respectively. The corresponding atomic radiative
time is about $T_r=3\times 10^{-2}s$. The transition $\left| g\right\rangle
\longleftrightarrow \left| e\right\rangle $ is strongly coupled to the
cavity mode with the coupling strength $\lambda =2\pi \times 25$ kHz. The
pump field can be provided by a classical microwave source. The adiabatic
approximation is valid under the condition that the time scale $T$ of the
variation of the control parameter is longer than the inverse of the energy
gap $\delta E=\lambda $ between the dark state and the nearest bright states
with nonzero eigenenergies, i.e., the dynamical time scale. If we set $%
T=20/\lambda $, then the leakage error to the excited eigenstates is on the
order of $1/(\lambda T)^2=2.5\times 10^{-3}$. Suppose that the Rabi
frequency $\Omega $ of the pump field is kept constant and the phase $\phi $
is slowly varied from $0$ to $2\pi $. Then due to dissipation the visibility
of interference fringes is approximately reduced by $(\Omega /\lambda
)^2T/T_c+\kappa _{g,f}T/2$, which is on the order of $10^{-3}$. The
correction to the geometric phase is $-2\pi (\Omega /\lambda )^2T/T_c$, also
on the order of $10^{-3}$. Therefore, the adiabatic condition can be
perfectly satisfied and the influence of decoherence is negligible. We note
that the atomic phase kick has been experimentally achieved by applying a
fast electric field pulse [32], which allows the measurement of the
geometric phase for the cavity field even if the system is not in the dark
state. In experiment, it may be more convenient to keep the phase of the
pump field unchanged, but detune its frequency from the frequency of the
quantized field by an amount $\delta $ with $\delta \ll \lambda $. Then $%
t\delta $ takes the role of the slowly varying phase $\phi $. An alternative
physical system to implement the required Hamiltonian is the circuit QED
setup, in which superconducting qubits are strongly coupled to the resonator
field. In fact, the Hamiltonian of Eq. (1) has been experimentally realized
using a phase qubit coupled to a superconducting coplanar waveguide
resonator and driven by an external microwave pulse [33].

The above ideas can be directly applied to the ion trap system. Consider an
ion trapped in a harmonic potential and driven by two laser beams tuned to
the carrier and the first lower vibrational sideband with respect to the
electronic transition $\left| g\right\rangle \rightarrow \left|
e\right\rangle $ and the vibrational mode. In the Lamb-Dicke limit, the
coupling between the internal and external degrees of freedom of the trapped
ions is described by the Hamiltonian (1), with the Rabi frequency and phase
of the laser for the carrier excitation serving as the control parameters.
The Berry phase acquired by the vibrational mode after an adiabatic
displacement evolution in phase space is directly manifested in the Ramsey
interference between $\left| g\right\rangle $ and an auxiliary state $\left|
f\right\rangle $ decoupled from the Hamiltonian. It is worthwhile to note
that a scheme has been proposed for measuring the geometric phase in the
vibrational mode of a trapped ion by adiabatically varying the squeezing
parameter in the engineered squeezing operator [34]. However, the scheme
requires squeezing transformations with opposite squeezing parameters before
and after the adiabatic evolution, discrimination between two nonorthogonal
coherent states, and techniques to cancel the accumulated dynamical phase.

We have shown how to produce and observe the nonclassical Berry phase of an
electromagnetic field in a generic qubit-boson system involving a qubit
coupled to a quantized field and driven by a classical pump field. The
adiabatic variation of the parameters of the pump field forces the quantized
field mode to displace along a loop in phase space, producing a purely
geometric phase. The origin of the geometric phase is the quantum nature of
the field mode without any classical counterpart. When the system is
initially in the dark state, the geometric phase can be directly detected
using the atomic Ramsey interferometer. Otherwise, one should apply a phase
kick to the atom to cancel dynamical contributions. Besides fundamental
interest, geometric manipulation of the quantized field opens new
possibilities for robust implementation of important quantum phase gates in
a single step. The effects of both the quantum and classical noises on the
Berry phase and visibility of the Ramsey interference fringes are analyzed.

Note added. Since completion of this work, two preprints by Vacanti et al.
[35] and Pechal et al. [36] investigating the measurement of the geometric
phase of a quantum harmonic oscillator using the interference of a qubit
have appeared. The first one proposed a scheme to measure the nonadiabatic
geometric phase produced by a coherent displacement force. The second one
reported an experiment for observing the Berry's adiabatic geometric phase
in circuit QED. In the experiment, the acquired phase includes a dynamical
contribution, and a purely geometric phase cannot be produced by the cyclic
evolution of the Hamiltonian. The geometric phase is only indirectly
measured by evaluating it as the difference between the phases for the
circular path and the phase for a straight line in the phase space. In
comparison, the present scheme allows direct measurement of this phase.

This work was supported by the National Fundamental Research Program Under
Grant No. 2012CB921601, National Natural Science Foundation of China under
Grant No. 10974028, the Doctoral Foundation of the Ministry of Education of
China under Grant No. 20093514110009, and the Natural Science Foundation of
Fujian Province under Grant No. 2009J06002.

\end{document}